\documentclass[]{spie}  %>>> use for US letter paper
\usepackage{xspace}
\newcommand{\harperfect}{\textsc{Harperfect}\xspace}
 
\usepackage{amsmath,amsfonts,amssymb}
\usepackage{graphicx}
\usepackage{siunitx}
\usepackage{aas_macros}
\DeclareSIUnit[]{\pix}{pix}
\usepackage{xcolor}

\usepackage[colorlinks=true, allcolors=blue]{hyperref}

\title{``Perfect'' spectra for ESO's HARPS spectrograph}

\author[a,b]{Dinko Milakovi{\'c}}
\author[a,b]{Guido Cupani}
\author[c,d]{Bruce A. Bassett}
\author[a,b]{Stefano Cristiani}
\author[e]{Luca Pasquini}
\affil[a]{INAF - Osservatorio Astronomico di Trieste, Via Tiepolo 11, Trieste, 34131, Italy}
\affil[b]{Institute for Fundamental Physics of the Universe, via Beirut 2, 34151 Trieste, Italy}
\affil[c]{School of Computer Science and Applied Mathematics, University of the Witwatersrand, Johannesburg, South Africa}
\affil[d]{Wits MIND Institute, University of the Witwatersrand, Johannesburg, South Africa}
\affil[e]{INAF – Osservatorio Astrofisico di Arcetri, Largo E. Fermi 5, 50125, Firenze, Italy}

\authorinfo{Further author information: (Send correspondence to D.M)\\D.M.: E-mail: dinko@milakovic.net, \\  G.C.: E-mail: guido.cupani@inaf.it}

\begin{document} 
\maketitle

\begin{abstract}
\harperfect\ is the first application of spectral perfectionism to a high-resolution echelle spectrograph, reconstructing 1D spectra from 2D detector data via a forward model built on a calibration matrix of PSF shape, wavelength, and order trace, built entirely from laser frequency comb (LFC) observations. HARPS cannot resolve individual LFC lines, so each of 21\,148 identified lines directly samples the PSF. We modelled these with a bivariate Gaussian, determining the PSF anywhere on the detector, calibrating wavelength, and tracing orders more accurately than the standard HARPS pipeline. Applied to 52.5 hours of HARPS data on quasar HE0515$-$4414, \harperfect\ gives S/N comparable to, though slightly lower than, the standard pipeline. Its value lies in an exact resolution matrix and independent samples, properties standard extraction cannot guarantee for a non-separable PSF -- a step toward better measurements of fundamental constants and exoplanet atmospheres.

\end{abstract}


\section{INTRODUCTION}
\label{sec:intro}  

The analysis of astronomical spectra starts with ``spectral extraction'', i.e.\ the conversion of the information recorded on two-dimensional digital detectors into a wavelength calibrated and flux calibrated one-dimensional data, a representation of the true input spectrum. The ``optimal'' method for the extraction of spectra from digital detectors was developed in the 1980s with the introduction of charge-coupled devices (CCDs) to astronomical spectrographs \cite{Robertson1986PASP...98.1220R, Horne1986PASP...98..609H}. The extraction process was since improved to compensate for curvature of spectral traces (particularly important for echelle spectrographs) and to improve the handling of the non-uniform response function of detector pixels \cite{Piskunov2002A&A...385.1095P, Zechmeister2014A&A...561A..59Z, Piskunov2021A&A...646A..32P}. A major simplifying assumption in optimal extraction is that the spectral traces are well aligned with the detector grid (columns and rows) and, consequently, that the point-spread function (PSF) of the spectrograph is a separable function of column $x$ and row $y$ \cite{Bolton2010PASP..122..248B}. This assumption is rarely satisfied in practice and, unless specific steps are taken to correct for this, reconstructed spectra contain errors in flux estimates, wavelength calibration, and a degraded resolution. Such effects are particularly damaging for studies which rely on accurate measurements of line shapes and line centres, e.g.\ studies of fundamental physical constants, isotopic abundances, and atmospheres of exoplanets \cite{Pasquini2025JATIS..11a1207P}. 

In this paper, we present the the first application of an advanced spectral extraction procedure, called ``spectral perfectionism'' (SP) \cite{Bolton2010PASP..122..248B}, to observations made by an extremely stable, high-resolution echelle spectrograph, HARPS \cite{Mayor2003Msngr.114...20M}. Spectral perfectionism was shown to extract spectra to the statistical noise limit for an arbitrarily complicated PSF, without making assumptions about its independence in $x$ and $y$ components, and in the presence arbitrarily high and wavelength-varying foreground \cite{Bolton2010PASP..122..248B}. However, this procedure requires a series of detailed information on the instrument (its calibration), as we briefly explain below. The resulting spectra are free from instrumental effects and are the closest possible representation of the incoming light -- provided that the calibration is correct. Our motivation for this work was to assess whether spectral perfectionism offers tangible benefits for extremely precise measurements. We chose HARPS because it is equipped with a laser frequency comb (LFC), which we used to study the spatial variations of the PSF and to calibrate the spectrograph. 

\section{SPECTRAL PERFECTIONISM FRAMEWORK}
\label{sec:framework}

Spectral perfectionism reconstructs the intrinsic object's spectrum, represented by a vector $\mathbf{f}$, by combining the observed spectrum (vector $\mathbf{p}$) with the information contained in a ``calibration matrix'' $\mathbf{A}$. This matrix is the collection of all available knowledge about the instrument, including the wavelength calibration, pixel response function, and the PSF shape as a function of detector position and wavelength \cite{Bolton2010PASP..122..248B}.

Mathematically, the forward model is expressed as $\mathbf{p}=\mathbf{A}\mathbf{f}+\mathbf{n}$, where $\mathbf{n}$ is the pixel noise vector. Assuming normally distributed noise, the unbiased estimator for the input spectrum can be found via linear least-squares:
\begin{equation}
    \mathbf{\hat{f}} = (\mathbf{A}^\mathrm{T}\mathbf{N}^{-1}\mathbf{A})^{-1}\mathbf{A}^\mathrm{T}\mathbf{N}^{-1}\mathbf{p},
\end{equation}
where $\mathbf{N}$ is the noise covariance matrix, defined as $N_{ij}=\langle n_i n_j\rangle$. Because the CCD pixels are assumed to have independent Poisson or Gaussian noise, $\mathbf{N}$ is strictly diagonal. 

The covariance matrix of this extracted spectrum is given by $\mathbf{C} = (\mathbf{A}^\mathrm{T}\mathbf{N}^{-1}\mathbf{A})^{-1}$. Because each resolution element illuminates several detector pixels, adjacent wavelength bins share illuminated pixels through PSF overlap, making the columns of $\mathbf{A}$ mutually non-orthogonal. As a result, $\textbf{C}$ has large off-diagonal elements and $\mathbf{f}$ suffers from extreme anti-correlated noise, known as covariance ringing. To suppress this noise, the spectrum must be reconvolved with a resolution matrix $\mathbf{R}$, yielding a smoothed, physically meaningful spectrum $\mathbf{\tilde{f}} = \mathbf{R}\mathbf{\hat{f}}$. The central challenge is choosing $\mathbf{R}$ so that this reconvolution does not degrade the native instrumental resolution. In the SP framework, the analogous choice is $\mathbf{R} \propto \mathbf{C}^{-1} = \mathbf{A}^\mathrm{T}\mathbf{N}^{-1}\mathbf{A}$, which yields a reconvolved spectrum $\tilde{\mathbf{f}} \propto \mathbf{A}^\mathrm{T}\mathbf{N}^{-1}\mathbf{p}$, a global weighted sum of detector pixels with weights proportional to the PSF divided by the variance. Because $\mathbf{R}\propto\mathbf{A}^\mathrm{T}\mathbf{N}^{-1}\mathbf{A}$, the effective kernel applied to the spectrum is the noise-weighted auto-correlation of the line-spread function (LSF). For a Gaussian PSF of width $\sigma$, this auto-correlation is a Gaussian of width $\sigma\sqrt{2}$, broadening the extracted LSF by a factor of $\sqrt{2}$ and degrading the native resolving power by nearly 30\%. To avoid this resolution penalty, Ref.\,\citenum{Bolton2010PASP..122..248B} proposed deriving $\mathbf{R}$ from the symmetric matrix square root of $\mathbf{C}^{-1}$ instead. We compute an intermediate matrix $\mathbf{Q}$ defined by:

\begin{equation}
\mathbf{C}^{-1} = \mathbf{Q}^2, \quad \mathbf{Q} = (\textbf{C}^{-1})^{\frac{1}{2}}, \quad \mathbf{Q} = \mathbf{Q}^T
%\mathbf{C}^{-1} = \mathbf{Q}^\mathrm{T}\mathbf{Q} \quad \implies \quad \mathbf{Q} = (\mathbf{C}^{-1})^{1/2}
\end{equation}

The final resolution matrix $\textbf{R}$ is constructed by row-normalising $\mathbf{Q}$ to ensure strict flux conservation. Writing $s_i = \sum_j Q_{ij}$ for the row sums used in this normalisation, the reconvolved spectrum has an exactly diagonal covariance matrix, with uncertainty $\sigma_i = 1/s_i$ on each element $\tilde{f}_i$. The rows of $\mathbf{R}$ represent the unbroadened LSF at each wavelength bin, preserving the native resolving power of the spectrograph to within the accuracy of the PSF model encoded in $\mathbf{A}$.

\section{CONSTRUCTING THE CALIBRATION MATRIX}
\label{sec:calib_matrix}

The calibration matrix contains all available knowledge of the instrument: the positions of illuminated pixels (the order trace), their central wavelength (wavelength calibration), pixel responses (flat-fields), as well as the PSF shape at each pixel. The most novel part of this work relates to using observations of HARPS's LFC to construct the calibration matrix.

The HARPS LFC is described in a series of articles, e.g.\ Refs.\ \citenum{Probst2014SPIE.9147E..1CP, Probst2020NatAs...4..603P, Milakovic2020MNRAS.493.3997M}. Briefly, it produces $\approx10\,000$ modes with frequencies given by the formula $f_n = f_0 + nf_{rep}$, where $n$ is the mode ordinal number (an integer), $f_{rep} = \qty{18}{\giga\hertz}$, and $f_0=\qty{4.58}{\giga\hertz}$ is the offset frequency of the first mode. The intrinsic width of one LFC line is $\sigma_f=\qty{20}{\kilo\hertz}$\cite{Milakovic2020MNRAS.493.3997M}, whereas the full-width at half maximum (FWHM) of the HARPS resolution element ranges from \qtyrange{4}{5.5}{\giga\hertz}, meaning that LFC lines are unresolved by HARPS. Consequently, the observed shape of a single LFC line is an excellent representation of the pixel-integrated PSF at that one detector position and wavelength. The idea was to map out how the LFC line shape vary across the detector, and use their known frequencies and positions to wavelength calibrate the spectrograph and trace echelle order centres.

\begin{figure}
    \centering
    \includegraphics[width=\linewidth]{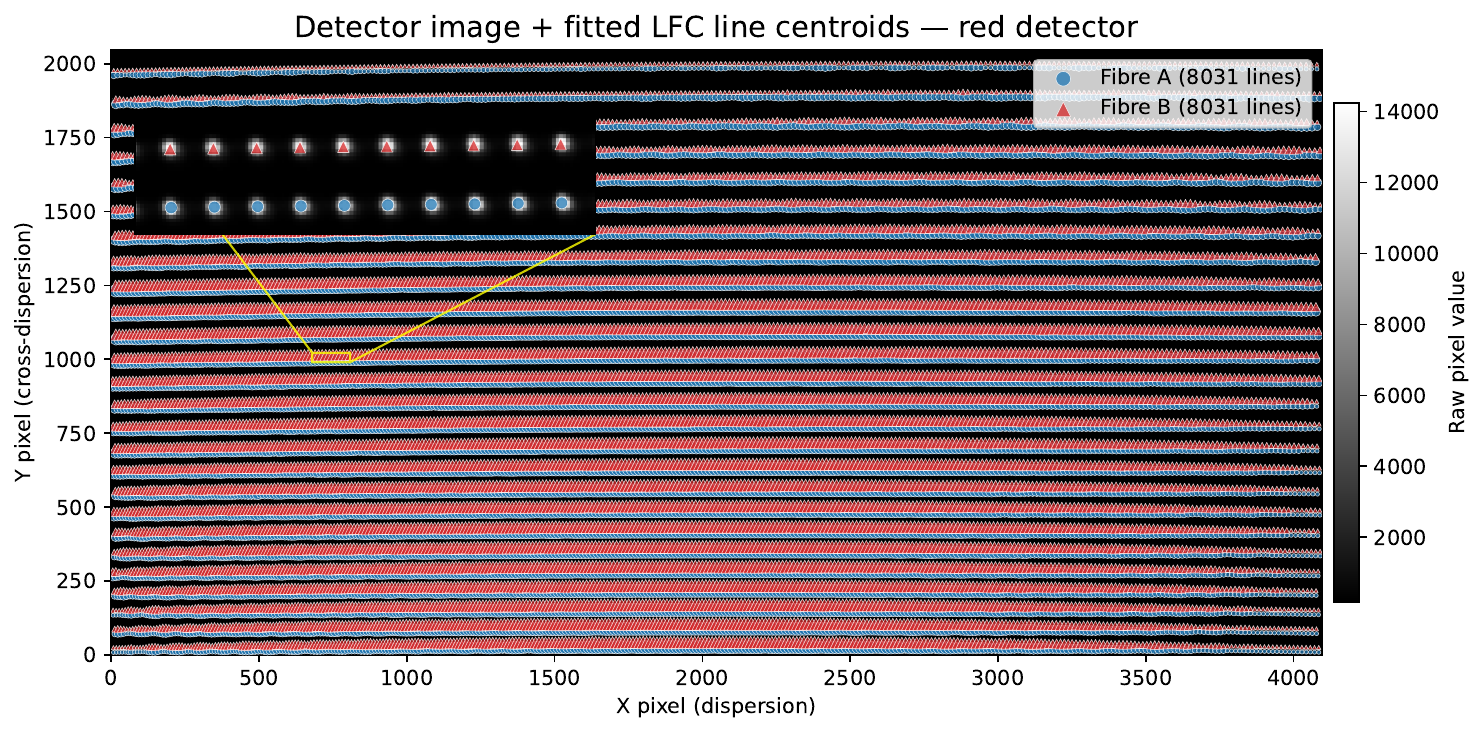}
    \caption{Locations of identified LFC lines with raw detector counts in the background (red detector). Blue dots and red squares indicate the centres of LFC lines identified in fibres A and B, respectively, used subsequently for wavelength calibration and order tracing. The inset shows a 128 pixel wide and 32 pixel high rectangle containing 10 LFC lines per fibre.}
    \label{fig:line_detection_red}
\end{figure}

We used a calibration exposure with LFC light feeding both HARPS fibres (referred to as fibres A and B), taken as a part of an ESO observing programme 0102.A-0697(A) \cite{Milakovic2021MNRAS.500....1M}. The HARPS detector is a mosaic of two CCDs, covering diffraction orders 89 through 114 (red detector) and 116 through 161 (blue detector), which we treated separately. Each CCD consist of $4096 \times \qty{2148}{\pix\squared}$ (in $x$ and $y$ directions, respectively), with the first and final 50 pixels in the $y$ direction being pre-scan and over-scan regions (and were hence removed from further consideration, leaving a $4096 \times \qty{2048}{\pix\squared}$ frame to work with). Running a maximum detection algorithm\footnote{We used \texttt{peak\_local\_max} from \textsc{scikit-image} package \cite{vanderWalt2014PeerJ...2..453V}.}, we identified 5086 LFC lines falling on the blue detector (2729 appearing on fibre A trace and 2357 on fibre B trace) and 16062 falling on the red detector (8031 per fibre). Organising the lines into orders based on their proximity along the main dispersion direction, we retrieved full spectral ranges for orders 89 through 123, covering the wavelength range $\qty{499.55}{\nano\metre} \leq \lambda \leq \qty{691.45}{\nano\metre}$. LFC lines identified on the red detector are shown on Fig.\,\ref{fig:line_detection_red}.

\subsection{The Point-spread Function, Wavelength Calibration, and Order Tracing}

We approximated the HARPS PSF shape with a bivariate (2-dimensional or 2D) Gaussian function. All 21148 LFC lines were independently fitted using this bivariate Gaussian function, with seven free parameters describing one line:  amplitude $A$, centre ($\mu_x, \mu_y$), standard deviation ($\sigma_x, \sigma_y$), rotation angle with respect to the $y$ axis ($\theta$, also used to measure the correlation between the elongation of the PSF in $x$ and $y$ directions), and a zero-offset. We optimised model parameters using the Trust Region Reflective algorithm \cite{Branch1999SJSC...21....1B}\footnote{As implemented in the \texttt{least\_squares} function of a python package \textsc{scipy}\cite{2020SciPy-NMeth}}. The region considered was an $11\times\qty{11}{\pix\squared}$ box centred on the brightest pixel. During parameter optimisation, we compared raw detector counts to the integrated flux under the Gaussian profile over the corresponding pixel area. For comparison, simply evaluating the Gaussian function at pixel centres resulted in 20\% larger average values for $\sigma_x$ and $\sigma_y$. We recorded the best-fit parameter values and their uncertainties in a FITS table for later use. 

\begin{figure}
    \centering
    \includegraphics[width=0.8\linewidth]{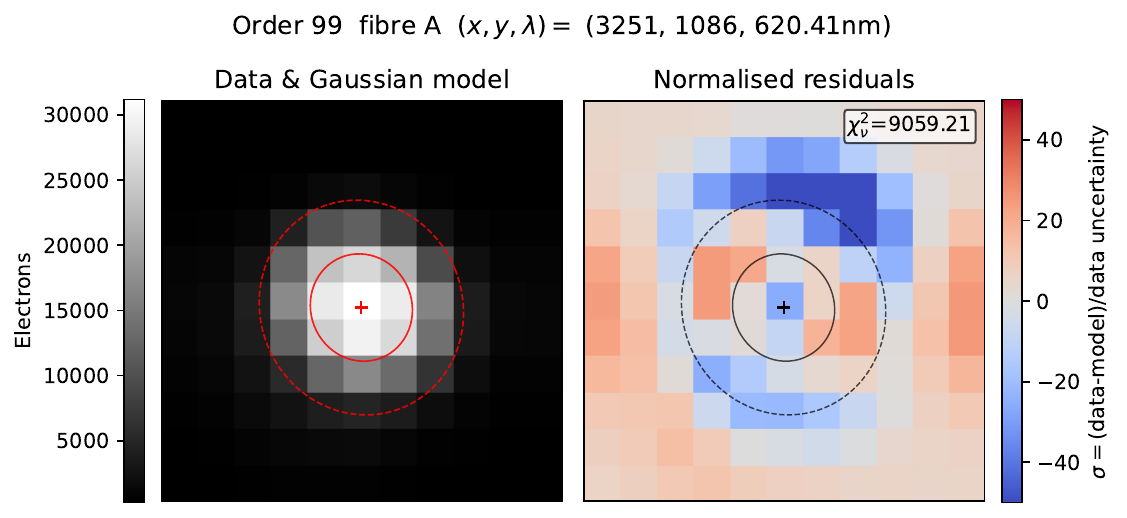}
    \caption{Randomly selected LFC line on the red detector. Left: raw electron counts together with the centre (red cross) and $1\sigma$ and $2\sigma$ contours (red lines) of the best-fitting 2-dimensional Gaussian model. Right: residuals normalised by the statistical uncertainty on the data (including read-out noise). The large $\chi^2_\nu$ value (printed in the top right) is not surprising for reasons explained in the text. In the right panel, red indicates pixels where $\mathrm{data}>\mathrm{model}$ and blue indicates the opposite. }
    \label{fig:line_fit}
\end{figure}

Left panel of Fig.\,\ref{fig:line_fit} shows the raw data and the best-fit Gaussian model for a randomly selected LFC line, with the right panel showing the corresponding normalised residuals: $(\mathrm{data}- \mathrm{model})/\sqrt{\mathrm{Var(data)}}$, where model is our bivariate Gaussian at best-fit solution and $\mathrm{Var(data)}$ is the variance on the data (all values are in electron counts). Some pixels have normalised residuals as large as $40\sigma$, bringing the reduced $\chi^2$ statistic of the fit to $\chi^2/\nu=9059$ ($\nu=114$, the number of degrees of freedom). Such high values are typical for all of our lines, but are not surprising, for the following three reasons. Firstly, variance on low-flux pixels is significantly smaller than the variance of high-flux pixels, resulting in unrealistically large residuals for the former. Secondly, besides including a zero-offset parameter, we did not model the spatially varying background light produced by non-linear processes in the photonic crystal fibre of the LFC system \cite{Probst2015PhDT.......270P, Probst2020NatAs...4..603P, Milakovic2020MNRAS.493.3997M}. This background follows the order trace and, in the residual plot, can be seen going horizontally through the centre of the plot. Thirdly, spatial correlations in the residuals demonstrate that a bivariate Gaussian is insufficient to capture the full PSF shape and its asymmetry. The asymmetry of HARPS's PSF is well documented from previous analyses \cite{Zhao2021A&A...645A..23Z, Milakovic2024A&A...684A..38M}.

\begin{figure}
    \centering
    \includegraphics[width=0.8\linewidth]{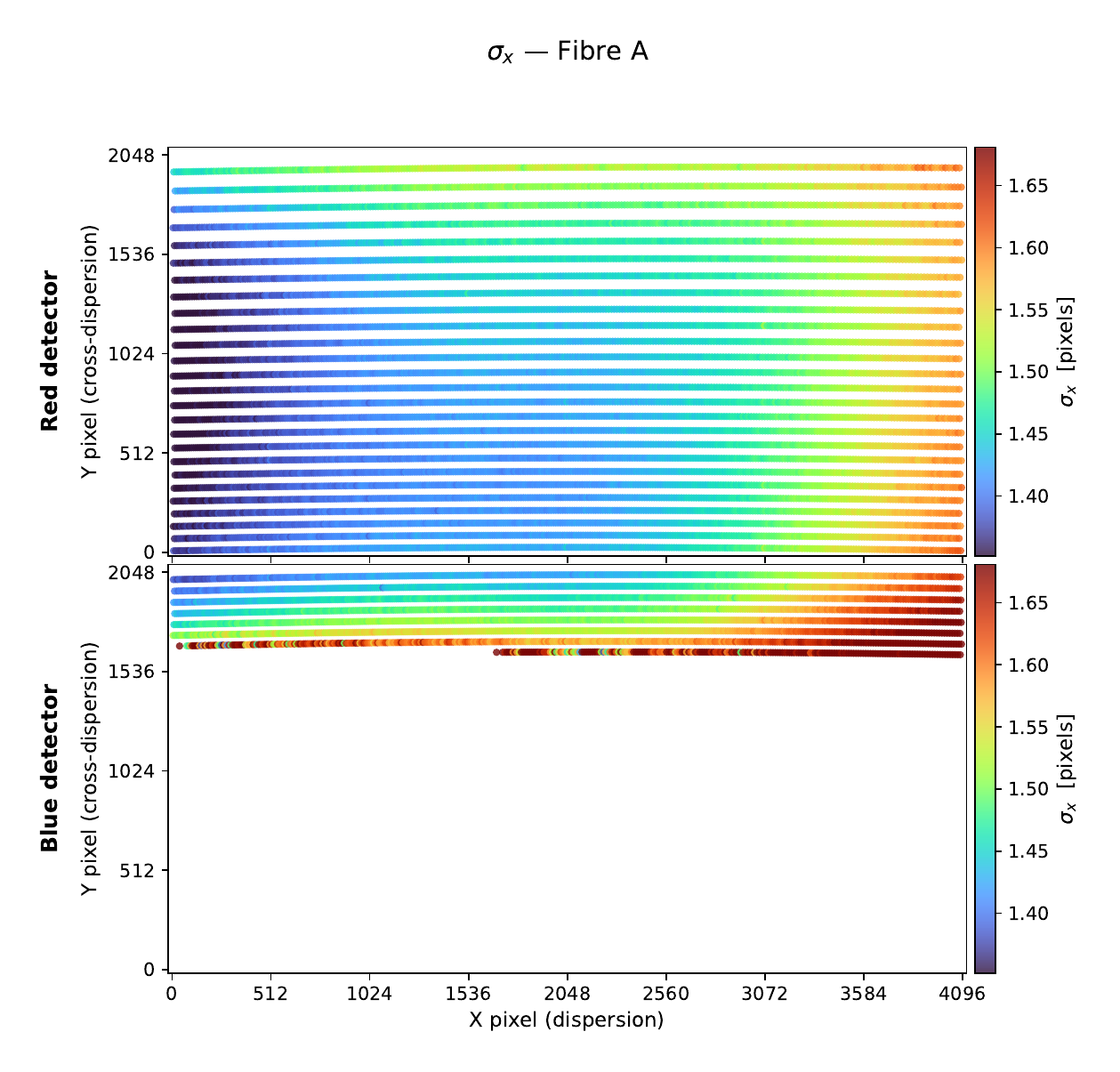}
    \caption{Standard deviation of the best fit 2-dimensional Gaussian function to individual LFC lines, projected along the detector's $x$ axis ($\sigma_x$), for fibre A. Each dot is a single LFC line we identified on both detectors, with colours indicating their $\sigma_x$ as per the colour bar to the right. The value of $\sigma_x$ is a proxy for spectral resolving power, and varies smoothly across the detector plane in a correlated manner, consistently across both detectors, indicating connection with instrument optics. In the panels, wavelength increases to the right and to the top. }
    \label{fig:sigma_x_fibre_A}
\end{figure}

\begin{figure}
    \centering
    \includegraphics[width=0.8\linewidth]{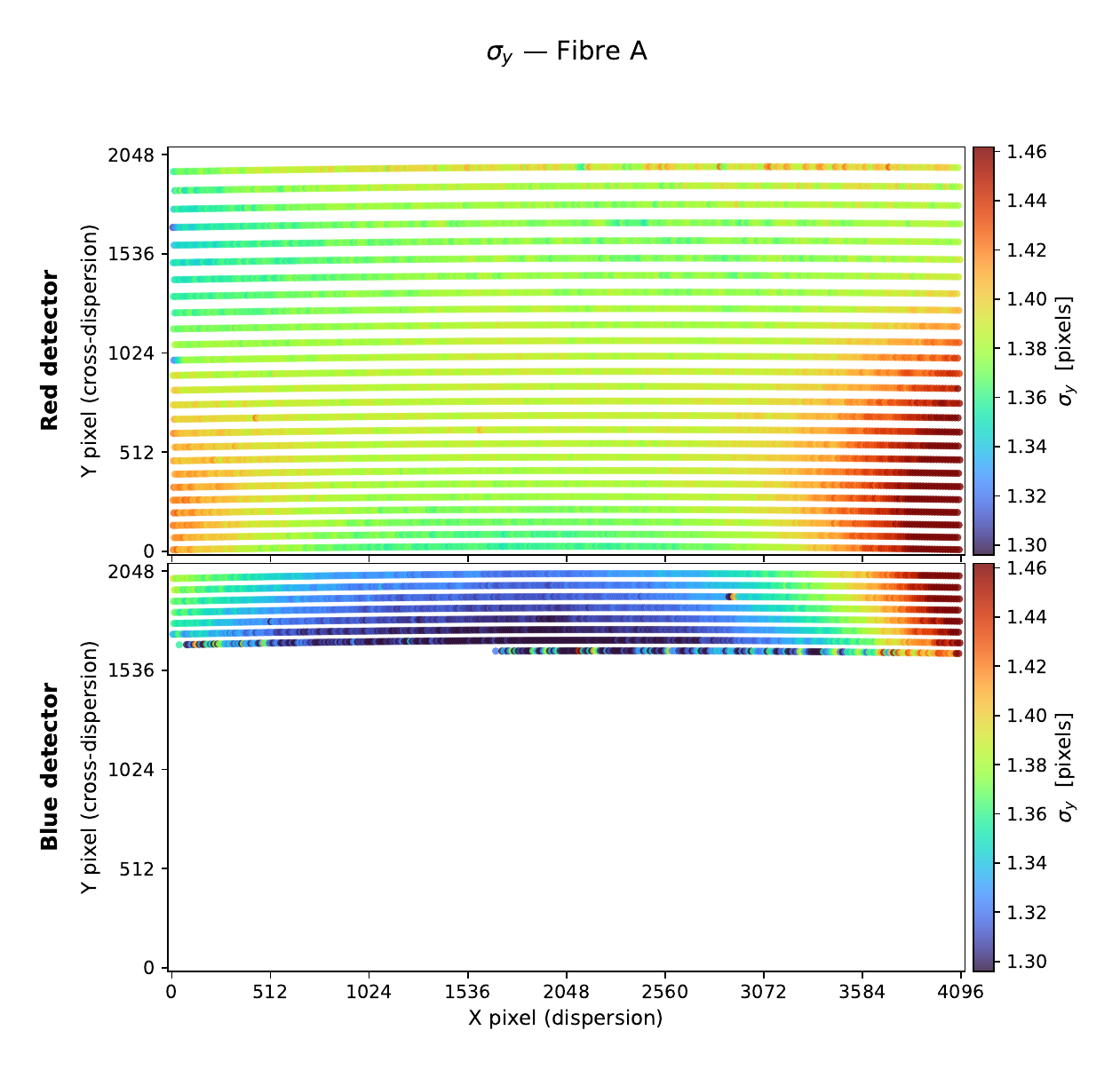}
    \caption{Similar as Fig.\,\ref{fig:sigma_x_fibre_A}, but for the projection of the PSF elongation along the $y$-axis ($\sigma_y$), tracing the size of the image projected by the optical fibre in the cross-dispersion direction. Also in this case, impacts of optical effects are clearly seen.     }
    \label{fig:sigma_y_fibre_A}
\end{figure}

We pause here to comment on the retrieved 2D Gaussian parameter values. Figures \ref{fig:sigma_x_fibre_A} and \ref{fig:sigma_y_fibre_A} show the values of $\sigma_x$ and $\sigma_y$ obtained from all LFC lines, with clear evidence of continuous spatial variation. This is not trivially expected because each of the plotted $\sim\num{10000}$ values (i.e.\ each dot in these figures is a single LFC line) were obtained independently, and there was no reason for best-fit parameter values to be correlated. These correlations are clearly seen in both the blue and the red detector, and there is a smooth change between the two detectors, consistent with them being caused by optical effects. Variations in $\sigma_x$ are consistent with anamorphism, i.e.\ the angle that the detector makes with the grating is larger in the blue end of each order than in the red, such that the resolving power varies within the order. The blue detector shows a curious, but incomplete, pattern due to the blue cut-off of the LFC wavelength range. Variations in $\sigma_y$ could hint to vertical astigmatism. Fibre B produced consistent results so is not shown. Observing such correlations from LFC line fits is a strong demonstration of the power that LFCs have in understanding how optical effects impact on instrument performance and for instrument diagnostics. It also supports the notion that we are indeed measuring PSF shape variations across the detectors.

\begin{figure}
    \centering
    \includegraphics[width=0.8\linewidth]{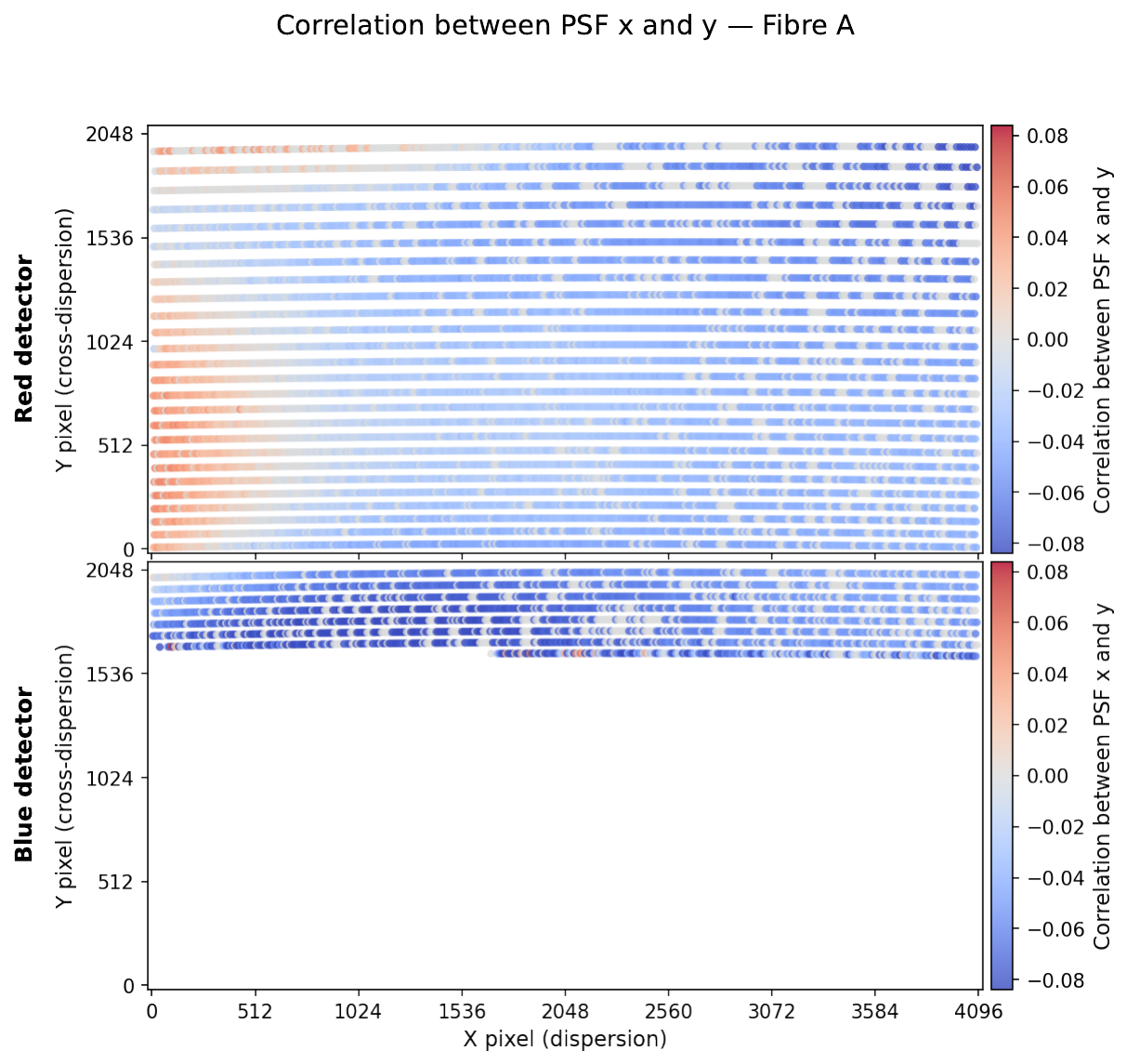}
    \caption{Correlation coefficient $\rho$ (Eq.\,\eqref{eq:rho}) between the $x$ and $y$ components of the bivariate Gaussian PSF model, fitted independently to each LFC line on fibre A, plotted as a function of detector position. $\rho=0$ corresponds to a PSF that is exactly separable in the detector $(x,y)$ frame and $|\rho|>0$ quantifies the degree of non-separability. Correlation coefficients vary smoothly across the cross the LFC wavelength range but remain modest ($|\rho|<0.1$ everywhere).}
    \label{fig:rho_fibre_A}
\end{figure}

Figure~\ref{fig:rho_fibre_A} shows the correlation coefficient $\rho$ between the $x$ and $y$ components of the bivariate Gaussian PSF, computed from the fitted $\sigma_x$, $\sigma_y$, and $\theta$ for each LFC line on fibre A. The coefficient $\rho$ is calculated using the following equations:
\begin{align}
    \rho &= \frac{\mathrm{Cov}(x,y)}{\sqrt{\mathrm{Var}(x)\,\mathrm{Var}(y)}}, \label{eq:rho} \\
    \mathrm{Cov}(x,y) &= \frac{1}{2}\left(\sigma_x^2 - \sigma_y^2\right)\sin(2\theta),\nonumber \\
    \mathrm{Var}(x) &= \sigma_x^2 \cos^2\theta + \sigma_y^2 \sin^2\theta, \nonumber \\
    \mathrm{Var}(y) &= \sigma_x^2 \sin^2\theta + \sigma_y^2 \cos^2\theta \nonumber.
\end{align}
The PSF is measurably non-separable across the full detector, but only weakly so, with $|\rho|$ remaining below $\sim0.08$ everywhere. On the red detector, $\rho$ varies smoothly and continuously, changing sign from positive at the bottom of the detector to negative toward the top. On the blue detector, $\rho$ is uniformly negative across nearly the entire fibre A trace. This pattern is consistent with the optical effects already inferred from the $\sigma_x$ and $\sigma_y$ maps (Figs.~\ref{fig:sigma_x_fibre_A} and~\ref{fig:sigma_y_fibre_A}), now expressed
directly in terms of the quantity that determines the validity of separable extraction.

The next step was to wavelength calibrate HARPS using the known LFC line frequencies. To ensure unambiguous identification of LFC line modes, we created a look-up table containing the $y$-coordinate and the wavelength of the central pixel in $x$ direction (pixel number = 2048) for all 72 echelle orders. The mode identification algorithm first selected all lines within one order, isolated the line whose $\mu_x$ was nearest to 2048, and then used the corresponding $\mu_y$ value to identify the diffraction order number from the look-up table. The mode of this line was calculated as the integer nearest to $n = \lfloor  (c/\lambda_\mathrm{cen} - f_0) / f_{rep}\rceil$, where $c$ is the speed of light and $\lambda_\mathrm{cen}$ is the wavelength of the central pixel from the look-up table. Following the identification of the central line, we assigned modes to all other lines stepping by one mode. 

\begin{figure}
    \centering
    \includegraphics[width=0.8\linewidth]{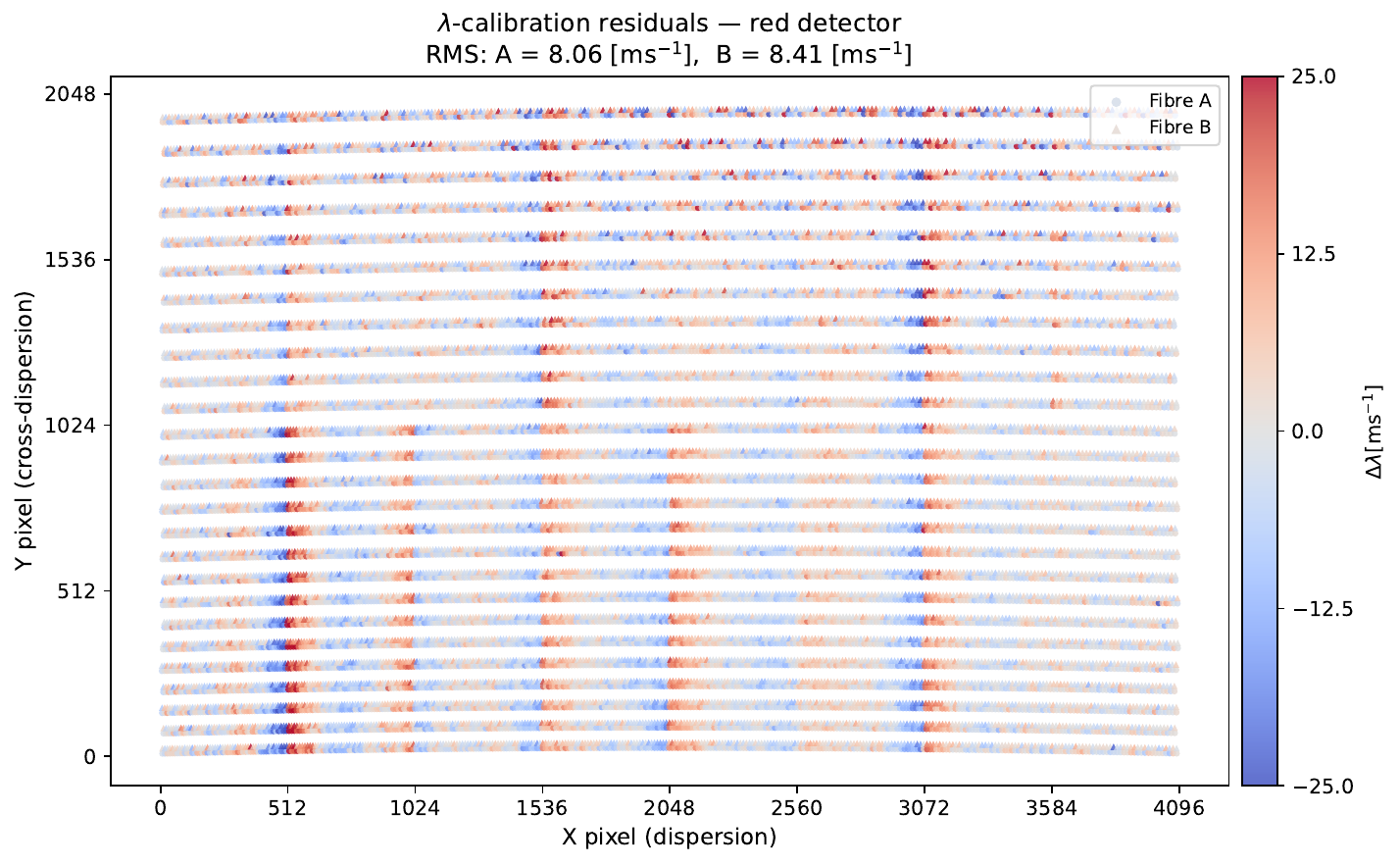}
    \caption{Residuals to the wavelength calibration as a function of position on the detector. Effects from uneven pixel sizes are clearly seen in the form of sharp $\pm\qty{25}{\metre\per\second}$ jumps every \qty{512}{\pix} in the $x$-direction.}
    \label{fig:residuals_map}
\end{figure}

We wavelength calibrated each order independently using the flexible Gaussian Process with a Matern $\nu=3/2$ kernel, appropriate for performing regression on once-differentiable physical functions. Hyperparameters of the kernel (signal amplitude and length scale) were optimised by maximising the log-marginal likelihood on the pairs of values ($\mu_x, \lambda)$, i.e.\ LFC line centre along $x$ and its wavelength. Derived Gaussian Process hyperparameter values are used later to obtain the wavelength of any given pixel. We confirmed the accurate identification of orders by comparing our wavelength calibration to an independent calibration produced by HARPS data reduction software (DRS) pipeline on ThAr arc lamp exposure taken on the same night, finding that the two match up to \qty{250}{\meter\per\second}. For comparison, a single HARPS pixel spans a wavelength range corresponding to $\approx\qty{820}{\meter\per\second}$, i.e.\ more than three times larger than the discrepancies between our LFC calibration and DRS's ThAr calibration. So far, we did not compensate for the uneven pixel sizes every \qty{512}{\pix} in the $x$ direction \cite{Wilken2012Natur.485..611W, Coffinet2019A&A...629A..27C, Milakovic2020MNRAS.493.3997M}, but we confirm that their effects are clearly seen in the wavelength calibration residuals (Fig.\,\ref{fig:residuals_map}). Compared to a normal distribution, the residuals are more centrally distributed, with zero mean and an root-mean-square of  \qty{8.06}{\meter\per\second} (\qty{8.41}{\meter\per\second}) for fibre A (B) on the red detector. 

\begin{figure}
    \centering
    \includegraphics[width=1\linewidth]{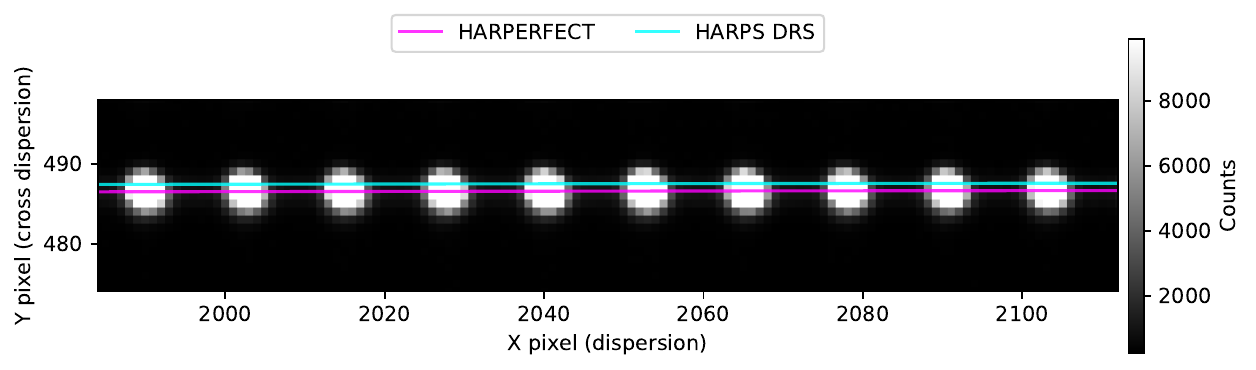}
    \caption{Section of the LFC spectrum showing differences between the order trace obtained by fitting a Gaussian Process to centres of LFC lines and the traces stored in HARPS DRS data products, derived from a tungsten lamp observation taken on the same night (\texttt{HARPS\_ORDER\_TABLE\_A.fits}). }
    \label{fig:order_trace}
\end{figure}

Finally, we used the same machinery to accurately trace out echelle order centres by performing Gaussian Process regression on pairs of ($\mu_x, \mu_y$) values. Interestingly, the centres derived by us using LFC lines seems to better trace the geometric centre of the order than the traces stored in data products of the HARPS DRS (see Fig.\,\ref{fig:order_trace}). The same behaviour is seen for all 34 orders with LFC coverage, on both detectors. The DRS determines order traces by fitting a 1-dimensional Gaussian to the cross-section of the trace (i.e.\ in the cross dispersion direction) at a fixed number of points along the order and fits a low order polynomial through them. We visually inspected other files, taken on the same night, to better understand the reasons for the discrepancy and identify which set of order traces -- ours or that of the DRS -- should be used. We found that traces derived from LFC line centres ($\mu_x,\mu_y$) almost always traced the geometric centre of the order better than the DRS trace. The only exception was in the two bluest orders illuminated by LFC light, where we did not have full coverage. The cause for the offset of the DRS traces is unknown, and the impact on scientific analysis is unquantified. 

\section{HARPERFECT}
\label{sec:workflow}

\subsection{Computational Feasibility and Sub-image Extraction}
Constructing a global calibration matrix $\mathbf{A}$ for the entire detector simultaneously would require working with a sub-image of $4096\times4096\approx17\,\mathrm{M}$ pixels, and a correspondingly sparse matrix $\mathbf{A}$ of dimensions $\sim17\,\mathrm{M}\times4096$, which is computationally prohibitive. Following previous implementations of the spectro-perfectionist approach (e.g., Ref.~\citenum{Guy2023}), we scale down the computational scope by processing localized segments of the detector rather than the entire frame simultaneously. Our implementation (\harperfect) concentrates on extracting a single fibre of one echelle order at a time, reducing the size of the matrix to be inverted. We defined a geometric bounding box, i.e.\ a narrow extraction ribbon of approximately \qty{60}{\pix} in height, containing the physical trace, and create a calibration matrix for that region of the detector. The height of the bounding box is determined empirically by extending the measured height of one order and padding it by \qty{20}{\pix} either side. This reduces the sub-image to a manageable $4096\times60$ pixels, and correspondingly $\mathbf{A}$ to $245{,}760\times4096$, allowing for a more rapid Cholesky decomposition, such that the extraction of the box takes approximately 50 seconds on a Macbook Pro from 2021.

\subsection{Simulating the PSF for an Arbitrary Position}
\begin{figure}
    \centering
    \includegraphics[width=1\linewidth]{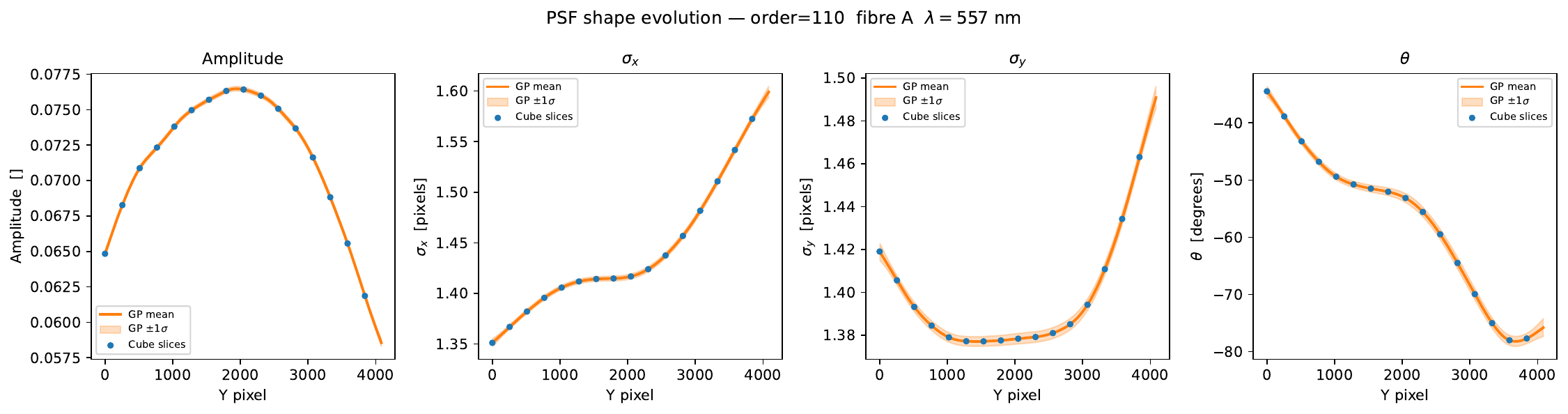}
    \caption{Evolution of the PSF shape parameters within diffraction order 110, fibre A. To create a mock PSF, we evaluate the regression function (orange line with uncertainty bands) at the desired $x$ coordinate and create a Gaussian PSF with the corresponding parameter values. Blue dots indicate the positions at which we created mock PSFs shown in Fig.\,\ref{fig:psf_gallery}.}
    \label{fig:psf_params_evolution}
\end{figure}

The SP approach demands an estimate of the wavelength and of the PSF shape for every extracted pixel. This becomes easy with the machinery we set up in Sec.\,\ref{sec:calib_matrix}. {\harperfect} contains methods to return the wavelength at an arbitrary position within the echelle order from the Gaussian Process regression described Sec.\,\ref{sec:calib_matrix}. Constructing the PSF shape for a specific pixel requires six parameters: the amplitude $\hat{A}$, the centre $(\hat\mu_x, \hat\mu_y)$, the standard deviations $(\hat\sigma_x, \hat\sigma_y)$, and the rotation angle $\hat\theta$. $\hat\mu_x$ is the centre of each of the 4096 physical pixels, and $\hat{\mu_y}$ is calculated from the previously determined order trace from $\hat\mu_x$. The values of the remaining four parameters ($\hat{A}, \hat\sigma_x, \hat\sigma_y, \hat\theta$) were obtained by interpolating between values saved in the FITS table containing the best-fit values of 2D Gaussian parameters fitted to LFC lines, together with their associated uncertainties, derived from procedures in Sec.\,\ref{sec:calib_matrix}. Also in this case, the interpolation was done using Gaussian Process regression to values ($x$, $p$), where $p$ is each of the four parameters. An example of the regression for order 110 is given in Fig.\,\ref{fig:psf_params_evolution}. 

\begin{figure}
    \centering
    \includegraphics[width=0.8\linewidth]{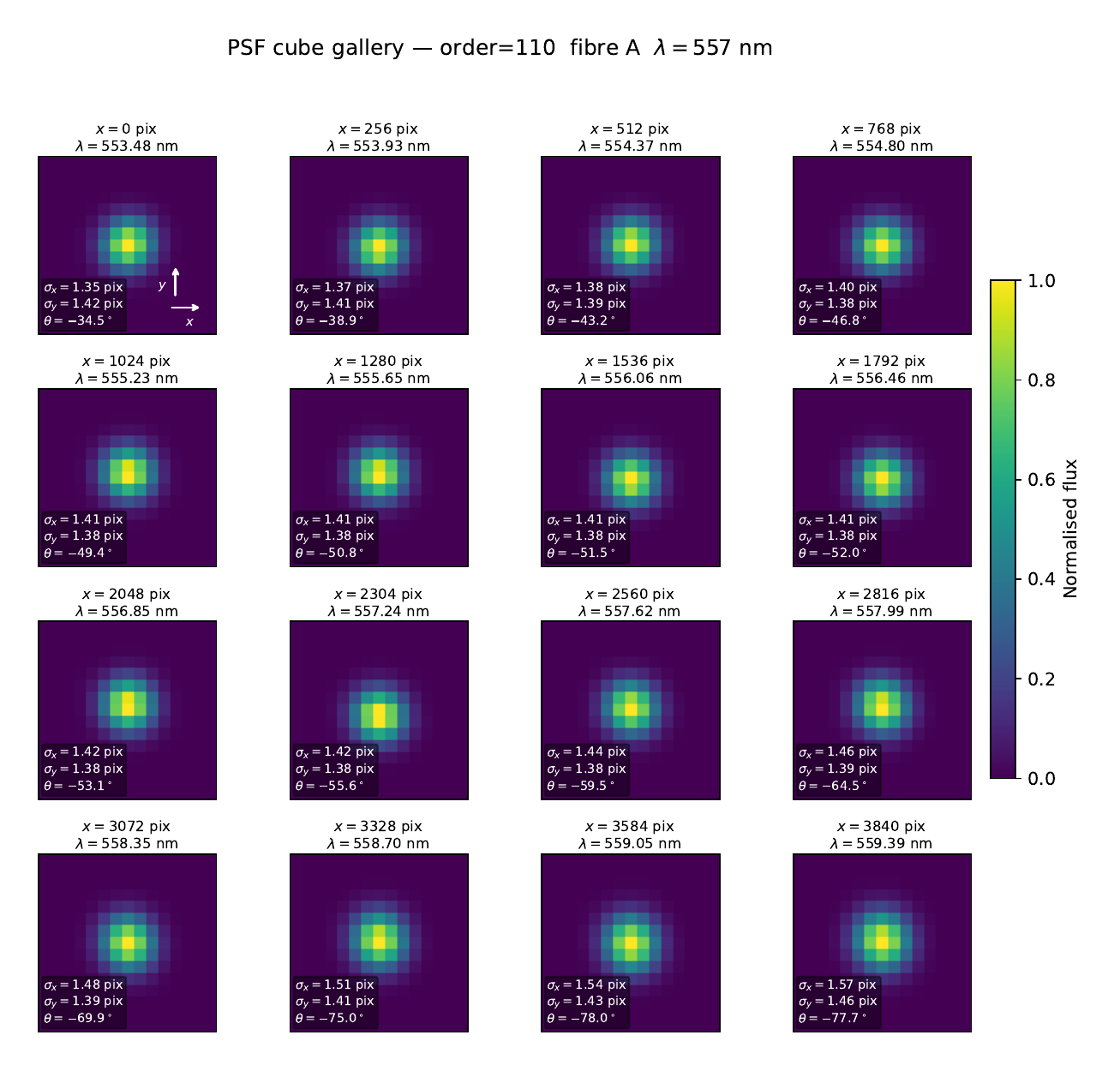}
    \caption{Gallery of simulated PSFs at 16 locations within diffraction order 110. Each panel shows the reconstructed PSF at a different position within the order (every 256 pixels). In the first panel, arrows indicate the direction of increasing $x$ and $y$. Text boxes report the values of $\hat\sigma_x$, $\hat\sigma_y$, and the rotation angle $\hat\theta$ of the PSF. }
    \label{fig:psf_gallery}
\end{figure}

In order to produce a PSF image for an arbitrary location along the recorded spectrum, the user is asked to specify the location of the FITS table containing fits to LFC lines, the ordinal number of the echelle order of interest (an integer between 0 and 71, for compatibility with DRS products), the target fibre (A or B) and the pixel number along the dispersion axis ($x$, a real number between 0 and 4095). The code automatically loads in the required information, performs Gaussian Process regression, saves intermediate products, and returns a wavelength and the PSF image. An example of the PSFs constructed in this manner are shown in Fig.\,\ref{fig:psf_gallery} for 16 equidistant positions along the echelle order (every 256 pixels).

\subsection{Workflow}
We implemented the Spectro-Perfectionism mathematics as a custom Python pipeline. In its current state, the pipeline processes one spectral trace at the time. Parameters provided by the user at input are:
\begin{itemize}
    \item path to the directory containing scientific frames to reduce;
    \item path to the FITS file with the output of the LFC analysis produced beforehand (see Sec.\, \ref{sec:calib_matrix}), used to construct $\textbf{A}$;
    \item the order and fibre (A or B) to extract. For compatibility with DRS products, the order indices run from 0 to 71;
\end{itemize}

The pipeline then processes all the science images within the directory through the following steps:
\begin{enumerate}
    \item \textbf{Data Ingestion:} For a given order/fibre combination, the pipeline loads the unextracted 2D science frame and queries the empirical LFC model to derive physical trace and crops a highly localized ribbon around it. At this time, we chose to use a pre-processed science frame, one of DRS intermediate products, i.e. a bias-subtracted, flat-fielded science exposures (using DRS nomenclature, {\harperfect} ingests \texttt{CCD\_corrected\_science.fits} files). The main reason was to isolate the effects of wavelength calibration, order tracing and PSF reconstruction on the final products while developing understanding, trusting that the bias subtraction and flat-fielding is correctly performed by the DRS. Coincidentally, this also reduces the size of the calibration matrix $\mathbf{A}$, speeding up calculations. Future versions of the code will improve on this aspect. 
    
    \item \textbf{Calibration matrix construction:} The pipeline defines its 1D extraction grid natively along the dispersion axis, corresponding one-to-one with the 4096 discrete physical columns of the detector. For each detector column $x$, the central wavelength and the physical $(x, y)$ centroid are evaluated directly from the empirical models derived from LFC lines. Concurrently, a 2D Gaussian PSF is generated (our model is already integrated over the pixel boundaries, Sec.~\ref{sec:calib_matrix}), populating the corresponding column of the sparse calibration matrix $\mathbf{A}$.
    
    \item \textbf{Linear inversion and deconvolution:} The sparse matrix $\mathbf{A}$ and the diagonal noise matrix $\mathbf{N}^{-1}$ are combined to compute the inverse covariance matrix $\mathbf{C}^{-1} = \mathbf{A}^\mathrm{T}\mathbf{N}^{-1}\mathbf{A}$. The raw, high-variance deconvolved flux vector $\mathbf{\hat{f}}$ is then solved via fast sparse Cholesky decomposition. 
    
    \item \textbf{Obtaining the final spectrum:} The pipeline computes the symmetric matrix square root of the dense inverse covariance ($\mathbf{Q} = (\mathbf{C}^{-1})^{1/2}$) and row-normalizes it to produce the flux-conserving resolution matrix $\mathbf{R}$. The raw deconvolved flux, $\mathbf{\hat{f}}$ from step 3, is smoothed by the resolution matrix to suppress covariance ringing, and to produce the final spectrum: $\mathbf{\tilde{f}} = \mathbf{R}\mathbf{\hat{f}}$, with a corresponding uncertainty spectrum derived from $\mathbf{Q}$. 
    
\end{enumerate}

\section{APPLICATION TO AN ASTRONOMICAL SPECTRUM}

In this section, we assess the impact of the SP approach on spectral quality of real astronomical data. We applied {\harperfect} to 36 HARPS observations of the quasar HE0515$-$4414, described in Ref.\,\citenum{Milakovic2021MNRAS.500....1M}, where they were used to put constraints on the variation of the fine-structure constant ($\alpha$) in a damped Lyman-$\alpha$ system at redshift $z\approx1.15$. The observations were taken in one week of December 2018, with a total exposure time of 52h 31m. The 36 new {\harperfect} spectra were all deblazed and corrected for Earth's motion with respect to the Solar system's barycentre. We applied the blaze function from a DRS product, \texttt{HARPS\_BLAZE\_A.fits}, made from tungsten lamp observations taken on the same night as quasar spectra. Barycentric Earth radial velocity (BERV) was read from standard DRS headers and applied to the wavelength array of {\harperfect} spectra. 

Finally, the spectra were combined in \textsc{Astrocook}\cite{2018SPIE10707E..23C,2020SPIE11452E..1UC}, using ``drizzling''-like technique on a wavelength grid with a bin size of \qty{0.830}{\kilo\meter\per\second}, matching the native pixel size of HARPS and the published DRS spectrum from Ref.\,\citenum{Milakovic2021MNRAS.500....1M}. With the drizzling approach, all contributions to each bin in the final grid are collected and weighted by their inverse variance and by the amount of superposition between the original pixel and the bin itself, achieving a statistically robust co-addition with a single rebinning procedure. In the last step, the continuum is derived by an iterative kappa-sigma clipping of absorption features and a subsequent Gaussian-kernel smoothing.

\begin{figure}
    \centering
    \includegraphics[width=\linewidth]{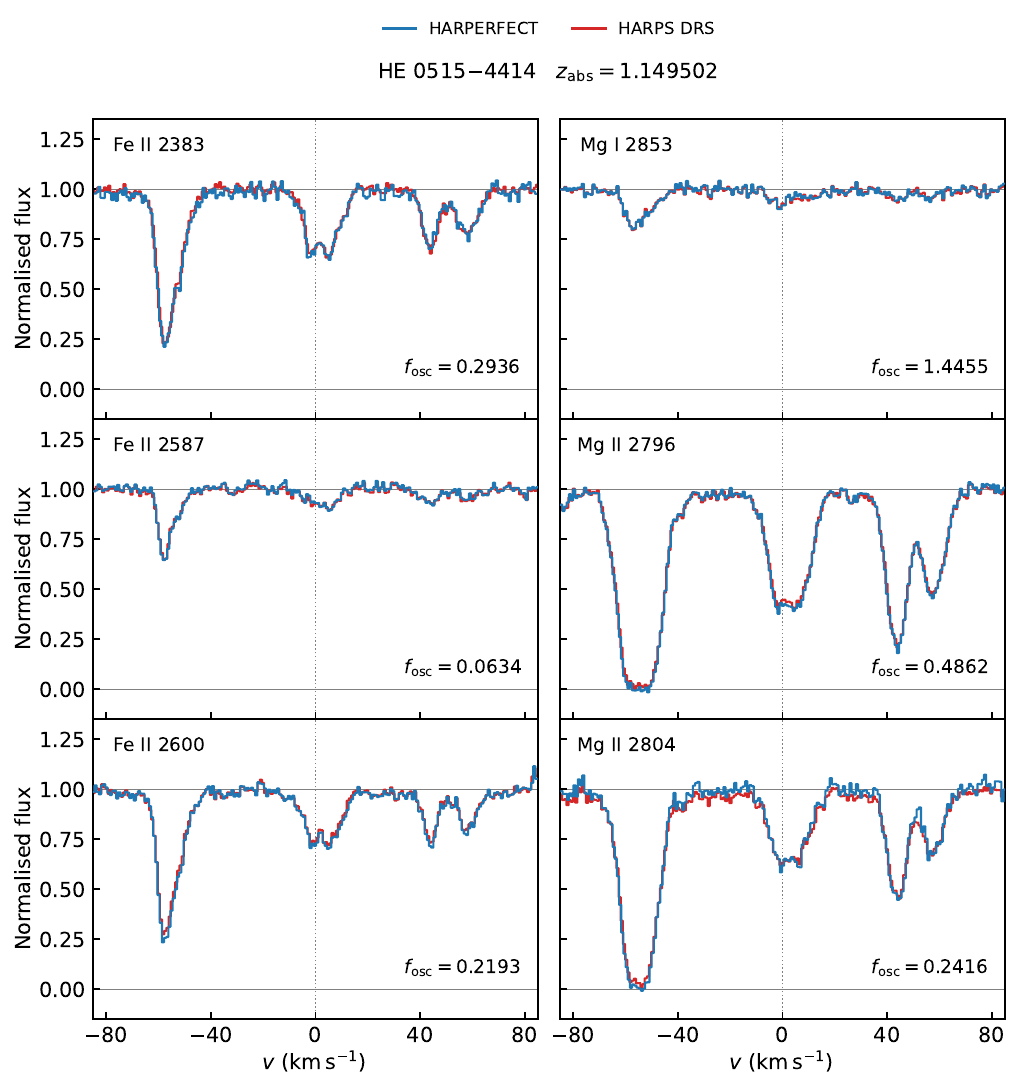}
    \caption{Small section of the absorption complex towards HE0515$-$4414, with zero velocity coinciding with the absorption at $z=1.149539$. Each panel shows a different transition, indicated in the top left, with the oscillator strengths in the bottom right. Two spectra are overplotted: the new spectrum produced by {\harperfect} (blue histogram) and a spectrum from the literature (red histogram), produced from the same observations using HARPS DRS \cite{Milakovic2021MNRAS.500....1M}. 
    }
    \label{fig:spectrum}
\end{figure}

Figure~\ref{fig:spectrum} compares a small section of the HE0515$-$4414 spectrum extracted by \harperfect\ to the corresponding HARPS DRS spectrum, for six metal absorption transitions at the same redshift. The two extractions agree closely for most of the transitions shown (Fe~\textsc{ii}~$\lambda2383$, Mg~\textsc{i}~$\lambda2853$, Fe~\textsc{ii}~$\lambda2587$, and Mg~\textsc{ii}~$\lambda2796$), with no visually discernible difference in line shape. For two transitions, Fe~\textsc{ii}~$\lambda2600$ and Mg~\textsc{ii}~$\lambda2804$, the \harperfect\ profile at $v=\qtylist{0;40}{\kilo\meter\per\second}$ appears visually narrower than the DRS profile near the line core. We have not yet fitted Voigt profiles for these components, so we report this as a qualitative visual impression rather than a measured effect, and we do not draw a general conclusion about resolution improvement from it. A quantitative comparison of fitted line parameters between the two extractions is left to future work.

\begin{table}[h]
\centering
\caption{Median S/N per \qty{1}{\kilo\metre\per\second} bin, \harperfect\ vs.\ HARPS DRS. The first column reports the wavelength range considered. The next two columns report values for {\harperfect} calculated in two different ways: the second column gives the S/N obtained by dividing the fluxes by their uncertainties obtained from the $\textbf{Q}$ matrix (method 1); whereas the third column provides S/N from dividing fluxes by the root-mean-square (RMS) of the fluctuations in the unabsorbed continuum (method 2). The fourt column S/N for HARPS DRS is derived from the flux variance array (method 1). The final column, $n$, reports the number of pixels used in calculating the median.}
\label{tab:snr_comparison}
\begin{tabular}{|c|c|c|c|c|}
\hline
Wavelength range (nm) & \harperfect (method 1) & \harperfect (method 2) & DRS (method 1) & $n$ \\
\hline
507.20 -- 510.40 & 62.29 & 64.88 & 65.84 & 2272 \\
560.30 -- 563.00 & 62.64 & 65.25 & 66.95 & 1737 \\
596.70 -- 600.00 & 55.27 & 57.58 & 60.95 & 1992 \\
609.36 -- 613.50 & 61.03 & 63.58 & 65.73 & 2446 \\
\hline
\end{tabular}
\end{table}

Table~\ref{tab:snr_comparison} reports the median signal-to-noise ratio per \qty{1}{\kilo\metre\per\second} velocity bin, computed in four representative wavelength windows free of all absorption, for both {\harperfect} and the standard HARPS DRS optimal extraction. The median S/N of the {\harperfect} extraction is consistently 1\% to 5\% lower than the S/N in the DRS extraction. To verify this, we calculated the numerical average of fluctuations in the unabsorbed continuum, and found it was more consistent with the DRS extraction. While initially surprising, this result can be understood as a consequence of the current modelling of the HARPS PSF. As a reminder, flux uncertainties are calculated by summing over rows of $\mathbf{Q}$, which is related to the calibration matrix $\textbf{A}$ containing PSF shapes. We know from Fig.\,\ref{fig:line_fit} that a Gaussian PSF is insufficient to fully describe HARPS's PSF and its asymmetry \cite{Zhao2021A&A...645A..23Z, Milakovic2024A&A...684A..38M}, meaning that $\textbf{A}$ is not strictly correct. Clearly, errors in $\textbf{A}$ have propagated into our uncertainty estimates, leading to lower S/N. Therefore, the result obtained here is consistent with -- rather than a failure of -- the framework.

\section{RESULTS}

{\harperfect} is the first application of spectral perfectionism to an extremely stable high-resolution astronomical spectrograph. In this first stage, the code is written with the HARPS spectrograph in mind, but is generally applicable to any other spectrograph. The most novel aspect of this work is the use of an LFC to construct the calibration matrix $\mathbf{A}$, a crucial component for the correct functioning of the SP framework. LFC lines were used to model the spectrograph's point-spread function, for wavelength calibration, and for order tracing. Our results can be summarised as follows:
\begin{enumerate}
    \item We modelled the PSF shape using a bivariate Gaussian function with seven free parameters. The machinery developed here allows mock PSFs to be created at an arbitrary position within the range covered by LFC light, i.e.\ in diffraction orders 89 -- 123\footnote{Not including order 115, which falls in the region between the two detectors, and is, hence, inaccessible.} (Fig.\,\ref{fig:psf_gallery}) by interpolating between measured Gaussian profile parameters. The Gaussian cannot fully capture the asymmetries present in the PSF shape (see Fig.\,\ref{fig:line_fit}) and is the immediate target for future improvement.

    \item We mapped out PSF shape variations across the detector with unprecedented detail, sampling it every $\approx15$ physical pixels (\qty{225}{\micro\meter}) along the dispersion direction. These variations are consistent with being caused by optical effects (Figs.\, \ref{fig:sigma_x_fibre_A}, \ref{fig:sigma_y_fibre_A}, and \ref{fig:rho_fibre_A}). The LFC has thus (again) proven to be a powerful tool for improving our understanding of the instrument physics. This opens the path to using the LFC for real-time instrument diagnostics and to understanding how instrument optics changes with time. In the future, the LFC could be used to study PSF variations due to thermal settling of instrument components, thermal breathing, and after interventions, among other applications. 

    \item Beyond $\sigma_x$ and $\sigma_y$ individually, we computed the correlation coefficient $\rho$ between the $x$ and $y$ components of the PSF, which quantifies its separability directly (Fig.~\ref{fig:rho_fibre_A}): $\rho=0$ corresponds to an exactly separable PSF, while $|\rho|>0$ indicates the degree to which the assumption underlying optimal extraction is violated. We find the HARPS PSF to be measurably non-separable across the full detector, but only weakly so, with $|\rho|$ remaining of order a few percent everywhere. This is consistent with the small but consistent S/N penalty of spectral perfectionism relative to optimal extraction reported below (item~5).
    
    \item Using the measured LFC line positions and their known wavelengths, we wavelength calibrated HARPS in two dimensions and traced the positions of diffraction orders with high accuracy. Our wavelength calibration matches the independent calibration performed by HARPS DRS using a ThAr lamp within \qty{250}{\meter\per\second}. Wavelength calibration residuals due to the known HARPS pixel size anomaly is approximately $\pm\qty{25}{\metre\per\second}$ and appears every \qty{512}{\pix} in the dispersion direction (Fig.\,\ref{fig:residuals_map}). Curiously, order traces  derived from LFC line centres seem to be more precise than the traces from a tungsten lamp produced by HARPS DRS, which always lays off-centre with respect to the visual geometric centre of the order (Fig.\,\ref{fig:order_trace}). The reason for this is unknown and the impact on science analysis derived from DRS products is unquantified.

    \item In a real-world application to the HARPS spectrum of the quasar HE0515$-$4414, the spectrum produced by {\harperfect} has a S/N that is a few percent lower than the spectrum produced by HARPS DRS (Tab.\,\ref{tab:snr_comparison}). This result should be viewed as a limitation of using a model that does not capture the true PSF shape, underlying the need to improve this aspect of the procedure to reap the full benefits of spectral perfectionism. 

    \item In a visual, unmodelled comparison of six metal absorption transitions at the same redshift (Fig.\,\ref{fig:spectrum}), four show no discernible difference between \harperfect\ and HARPS DRS, while two (Fe~\textsc{ii}~$\lambda2600$ and Mg~\textsc{ii}~$\lambda2804$) show apparently sharper profiles in the cores of narrow absorption components in {\harperfect} spectra. Given the absence of quantitative Voigt profile fitting, we present this as a preliminary visual observation rather than evidence of a systematic resolution gain, particularly in light of the S/N result above.

\end{enumerate}

Further improvements are expected by using a more accurate PSF model, modelling the LFC background, and compensating for the uneven pixel sizes during wavelength calibration. In its current form, {\harperfect} takes 50 seconds to processes a single box of $4096\times\qty{60}{\pix\squared}$, or 2 hours for the full spectrum. While this is substantially slower than optimal extraction, the increase in spectral fidelity -- required for precision measurements of absorption line centroids and shapes -- justifies the extra effort. 
    
\acknowledgments 

D.\,M.\, acknowledges the support provided by the Italian Institute of Astrophysics 2023 Fundamental Research Techno Grant ``Spectro-perfectionism for high-fidelity spectroscopy'' awarded to G.\,C. Based on observations made with ESO Telescopes at the La Silla Paranal Observatory under programme ID 0102.A-0697(A). D.\,M.\, thanks Paolo Molaro for interesting discussions that improved this work. 

Declaration of AI-Assisted Drafting and Development: Following COPE and SPIE guidelines, the authors acknowledge the use of Google Gemini and Anthropic's Claude as AI assistants in the writing and refinement of the underlying {\harperfect} codebase. The authors conceived the software architecture, thoroughly reviewed and validated all generated code, and retain full responsibility for the scientific and technical content herein.

% References
\bibliography{report} % bibliography data in report.bib
\bibliographystyle{spiebib} % makes bibtex use spiebib.bst

\end{document}